 \documentstyle[twoside,fleqn,espcrc2,epsfig]{article}

\def\lab#1      {\hbox{\small #1} }
\newcommand{\be}{\begin{eqnarray}}
\newcommand{\ee}{\end{eqnarray}}
\newcommand{\ben}{\begin{eqnarray*}}
\newcommand{\een}{\end{eqnarray*}}
\newcommand{\la}{\langle}
\newcommand{\ra}{\rangle}
\newcommand{\half}{\frac{1}{2}}
\newcommand{\pe}{\rightarrow}

\def\mb#1         {\mbox{\boldmath $#1$}}
\def\diffn#1	  {\Delta^{-}_{#1}}

\newcommand{\AmS}{{\protect\the\textfont2
  A\kern-.1667em\lower.5ex\hbox{M}\kern-.125emS}}

\begin{document}

\title{Connections between Tomboulis vortices and projection vortices}

\author{A. Alexandru and Richard W. Haymaker \\
{Department of Physics and Astronomy, Louisiana State University, Baton Rouge, LA 70803, USA}
\thanks{Presented by R. Haymaker. A. Alexandru's present address: Department of Physics, University of Colorado, Boulder, CO 80309, USA.  Supported in part by U.S. Dept.
                of Energy grant DE-FG05-91 ER 40617.}}

\begin{abstract}

By using the freedom of picking a representative we explore 
connections between the Tomboulis $SO(3) \times Z(2)$ form of the partition function
and the $SU(2)$ form.   We are able to express the  monopole and vortex observables of the former
in terms of configurations of the latter.  Also we can measure Tomboulis and projection
vortex counters on the same configuration to search for correlations.
\vspace{1pc}
\end{abstract}

\maketitle

\section{Introduction}

In 1980, Tomboulis derived an alternative form of the $SU(2)$ partition function  which 
is invariant under sign flips of the links\cite{t}.  This effectively maps the $SU(2)$
manifold to $SO(3)$. The $Z(2)$ dependence is carried by new plaquette valued variables.  By 
breaking down the $SU(2)$ group into its $SO(3)$ and $Z(2)$ factors, the underlying
monopoles and vortices are revealed.  This was further developed by Kovacs and Tomboulis\cite{kt1} [KT]
in which thin, thick and hybrid vortex linkages with a Wilson loop could be defined.

In a recent paper\cite{ah3} we made a more direct contact
between this approach and the usual SU(2) formalism. We made use of  $SO(3)$
{\em representative} invariance by making appropriate sign flips of the links.
We point out that there is a representative for which the two formulations are identical.
Therefore we can define the KT vortex counters on ordinary SU(2) configurations.

Secondly, by choosing another representative we show how projection vortices\cite{dfgo} arise naturally
in this formalism.  This also gives an alternative perspective on the projection approximation.

\section{$SU(2)$ configurations in $SO(3) \times Z(2)$ variables}

The Wilson form of the partition function can be recast by introducing
$Z(2)$ valued independent variables $\sigma(p)$ defined on 
plaquettes\cite{t,ah1}
\ben
Z_{SO(3) \times Z(2)}  &=& 
\int 
\left[
dU(b) 
\right]
\sum_{ \sigma(p)}
\\
&&
\left[
\prod_c
   \delta\left(\sigma(\partial c) \eta(\partial c)\right)
\right]
\\
&&
\exp
\left(
   \beta \sum_{p} \half|Tr[  U(\partial p)]| \sigma(p)
\right),
\een
where the dependent variables $\eta(p)$ are defined by
\ben
Tr [U(\partial p)] \equiv |Tr [U(\partial p)] | \eta(p).
\een
The ``cube constraint" factor requires that  $\prod_1^6 \eta(p)\sigma(p) = +1$ over the 
six faces of all cubes.

Wilson loops have  $Z(2)$ valued plaquette tiling factors, $\sigma$ and $\eta$ on an 
arbitrary surface $S$ bounded by $C$
\be
W_{m \times n}(C) &=& \lab{Tr} [  U(C)]  \eta(S) \sigma(S)|_{C = \partial S}, 
\label{wilson}
\ee
\ben
W_{1 \times 1} &=&  \lab{Tr} [  U(\partial p)] \eta(p)  \sigma(p) 
= |\lab{Tr} [  U(\partial p)] |  \sigma(p).
\een

Properties of this form include:
\begin{itemize} 

   \item$Z(2)$ invariance of $Z$ and of observables under   $U(b) \rightarrow - U(b)$. 
    There are therefore $2^N$ {\em representatives} of $SO(3)$, where $N$ is the number of links. \\

   \item There exist  co-closed $\sigma(p) - \eta(p)$ vortex sheets due to the cube constraint with 
        patches of either 
         $\sigma(p)=-1$ or $\eta(p)=-1$, $\sigma(p) \eta(p) = -1$. 
      Pure $\sigma(p)$ or $\eta(p)$  vortex sheets are limiting cases.\\
   \item A change of representative can deform existing $\eta$ patches and create or destroy pure
     $\eta$ vortex sheets.

\end{itemize}

\subsection{The representative $\widetilde{U}(b)$}

This is defined by the condition
\ben
\sigma(p)\eta(p) &=& +1,\;\;\;\forall p.
\een
In this case the cube constraint is automatically satisfied.
There are further simplifications:
\ben
|Tr[  \widetilde{U}(\partial p)]| \sigma(p) 
&=& Tr[  \widetilde{U}(\partial p)] \eta(p) \sigma(p),
             \\ &=& Tr[  \widetilde{U}(\partial p)], 
\een
\ben
\widetilde{Z}
&=& 
\int 
\left[
    d \widetilde{U}(b)
\right]
\exp
\left(
   \beta 
   \sum_{p} 
   \half\lab{Tr} [  \widetilde{U}(\partial p)] 
\right),
\een
\ben
W_{m \times n} = \lab{Tr} [  \widetilde{U}(C)] , \;\; 
  W_{1 \times 1} =  \lab{Tr} [  \widetilde{U}(\partial p)].
\een

We showed\cite{ah1,ah3} that starting from a cold configuration,
$U(b)=\sigma(p)=+1$, we can reach the full configuration space of the 
independent variables $\{U(b), \sigma(p)\}$ through local updates 
while staying in the representative $\widetilde{U}(b)$.  
In this representative all $\sigma-\eta$ vortices are absent.

This particular representative provides the connection of this formulation to the 
SU(2) formalism with the Wilson action.
As a consequence, we can define the Tomboulis thin, thick and hybrid
vortex counters on ordinary $SU(2)$ configurations as will be given below.

\subsection{The representative $\widehat{U}(b)$}

This is defined by the condition 
\ben
\lab{Tr} [\widehat{U}(b)] &\ge& 0.
\een
This can be obtained by a single sweep.
The interest in this is to connect with projection vortices which are defined 
as follows:  One first fixes the gauge, for example the maximal center gauge and then

{\bf In an arbitrary representative}
\begin{itemize} 
      \item Project: $ sign \lab{Tr} [U(b)] \rightarrow u(b)$, $u(b) = \pm 1$.
      \item $P$ vortex: $u(p) = u(\partial p) \eta(p) \sigma(p)=-1$
      \item Proj. approx.:  $W(C) \approx   u(S)|_{C = \partial S}$.
\end{itemize}

{\bf In the  $\widehat{U}(b)$ representative}
\begin{itemize} 
      \item Project: $ sign \lab{Tr} [\widehat{U}(b)] \rightarrow \widehat{u}(b)$,
       $\widehat{u}(b) = + 1$.
      \item $P$ vortex: $\widehat{u}(p) = \eta(p) \sigma(p)=-1$, which is identical to
$\sigma - \eta$ vortex.
        \item Proj. approx.:   $\lab{Tr} [\widehat{U}(C)] \approx 1 $,
\end{itemize}
where we have used Eqn.(\ref{wilson}).  These two procedures give identical P vortices. 

However in the $\widehat{U}(b)$ representative
the $P$ vortices are identical to the $\sigma - \eta$ vortices which are a
tiling factor in the exact definition of the Wilson loop.  The success or failure of a projection approximation 
depends on whether  one can find a gauge such that the sign fluctuations of the perimeter 
factor in Eqn.(\ref{wilson}) can be transferred to the tiling factors arising from $\sigma - \eta$ linkages.
If so then one argues that the area law of a Wilson loop arises
from P vortex linkages in that gauge.

\section{Kovacs-Tomboulis vortex counters}
Kovacs and Tomboulis\cite{kt1} gave representative independent definitions of three
vortex counters based on  $SO(3) \times Z(2)$  configurations.
\ben
N_{thin}(S)&=&\prod_{p\in S} \sigma(p), \\
N_{thick}(S)&=&\lab{sign} \; \lab{tr} [ U(C) ] \times \prod_{p\in S} \eta(p), \\
N_{hybrid}\;\;\;\;&=& N_{thin}(S) \times N_{thick}(S) = \lab{sign} \, W.
\een
The hybrid counter is necessarily independent of surface. 
$N_{thin}(S)$ and $N_{thick}(S)$ count the corresponding vortices only if the
value is independent of surface $S$.

We can express these counters in terms of $SU(2)$ configurations by evaluating the
above expressions in the $\widetilde{U}(b)$ representative.
\ben
N_{thin}(S)&=&\prod_{p\in S} \lab{tr} [\partial \widetilde{U}(p)], \\
N_{hybrid}\;\;\;\;&=& \lab{sign} \,\, \lab{tr} [\widetilde{U}(C)], \\
N_{thick}(S)&=& \prod_{p\in S} \lab{sign} \,\,\lab{tr} [\partial \widetilde{U}(p)]\times 
\lab{sign} \,\, \lab{tr} [\widetilde{U}(C)].
\een

\begin{figure}[h]
\begin{center}
\epsfig{file=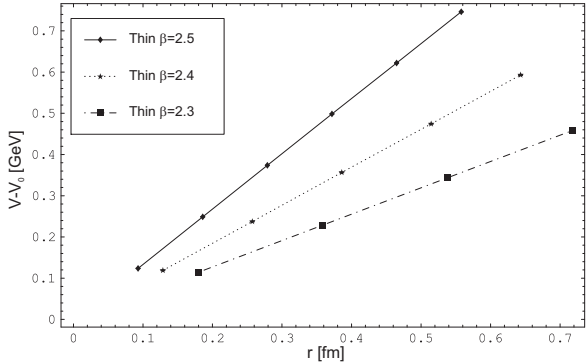, width=7.7cm}
\end{center}
\caption{Vortex potentials in physical units for the thin counter}
\end{figure}

\begin{figure}[h]
\begin{center}
\epsfig{file=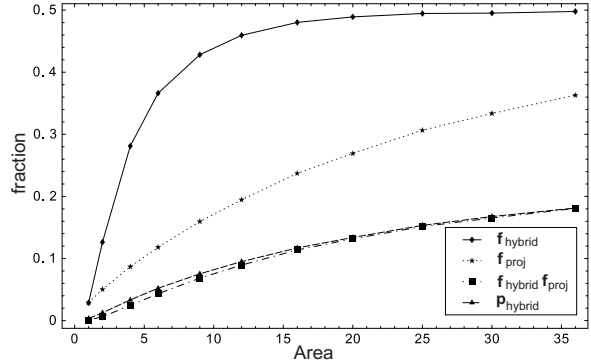, width=7.7cm}
\end{center}
\caption{Average of the fraction  odd/ (odd+even) hybrid and projection vortices linking a Wilson loop as 
a function of area.}
\end{figure}

\section{Numerical Results}

It is not feasible to measure these counters on all possible surfaces.  We made measurements only on the minimal
surface\cite{ah3,ah1}.  As a consequence, a measurement giving for example $N_{thin}(S) = -1$ indicates only the occurence
of an odd number of $\sigma$ patches which could be part of  thin or hybrid vortices. And similarly for
the thick case.

The contribution to the potential from the three types of vortex counters is
\ben
V(R)=-\lim_{T\pe\infty} \frac{1}{T} \ln \la N(W(R, T)) \ra,
\een
where $N(W(R,T))$ is the thin, thick or hybrid counter signal for that particular Wilson loop (taking values  $\pm 1$). 

Fig. 1 shows that the string tension in $V_{thin}$ in physical units {\em increases} in the continuum limit.
Although this is perhaps surprising, we showed that this is canceled by an increasing string tension in the thick potential\cite{ah3}.

The K-T definition for vortices\cite{kt1} is appealing since it is  gauge invariant 
 but they are hard to localize on a lattice. 
P vortices\cite{dfgo}, on the other hand, are easy to localize but are not gauge invariant. 
It is interesting to see if these two definitions agree. 
We now have the tools to compare these definitions of  vortex counters 
on the same configuration. 

Fig. 2 shows plots of the average of the fraction odd/(odd+even)  hybrid and projection vortices linking a Wilson loop as 
a function of area. The average of the product compared to the product of the average shows that there is essentially no correlation.  The corresponding plots for thin vortex fractions and thick ones gives essentially the same
result.  In Ref.\cite{ah3} we examine more sensitive signals of correlations but without a definitive result.


\begin{thebibliography}{99}
\bibitem{t} {E.T. Tomboulis, Phys. Rev. {\bf D23}, 2371 (1980).}
\bibitem{kt1} {T. G. Kovacs and E. T. Tomboulis, Phys. Rev. {\bf D} 57, 4054 (1998).}
\bibitem{ah3} {A. Alexandru and R.W. Haymaker, hep-lat/0108004 to be published in Physics Letters.}
\bibitem{dfgo}   { L. Del Debbio, M. Faber, J Greensite 
 and S. Olejnik, Phys. Rev. {\bf D} 55, 2298 (1997).}
\bibitem{ah1} {A. Alexandru and R.W. Haymaker, Phys. Rev. {\bf D62}, 074509 (2000).}
\end{thebibliography}
\end{document}